\documentclass[aps,prl,twocolumn,showpacs,superscriptaddress,groupedaddress]{revtex4}
\usepackage{graphicx}
\usepackage{hyperref}
\usepackage{amsmath}

\begin{document}

\title{Sub-cycle symmetry breaking of atomic bound states interacting with a short and strong laser pulse in a time-domain picture}

\author{Veit Stoo{\ss}$^{1}$, Andreas Kaldun$^1$, Christian Ott$^1$, Alexander Bl{\"a}ttermann$^1$, Thomas Ding$^1$\\ 
and Thomas Pfeifer$^{1,*}$\\
\small$^1$Max Planck Institute for Nuclear Physics, Saupfercheckweg 1, 69115 Heidelberg, Germany\\
\small$^*$Corresponding author: thomas.pfeifer@mpi-hd.mpg.de}
\date{\today}

\begin{abstract}In any atomic species, the spherically symmetric potential originating from the charged nucleus results in fundamental symmetry properties governing the structure of atomic states and transition rules between them. If atoms are exposed to external electric fields, these properties are modified giving rise to energy shifts such as the AC Stark-effect in varying fields \cite{Autler1955} and, contrary to this in a constant (DC) electric field for high enough field strengths, the breaking of the atomic symmetry which causes fundamental changes in the atom's properties. This has already been observed for atomic Rydberg states with high principal quantum numbers \cite{Haroche1977,Koch1984PRA,Vrakking2008,Vrakking2013PRL}. Here, we report on the observation of effects linked to symmetry breaking in helium within the optical cycle of the utilized strong visible laser fields for states with principal quantum number n=2. These findings were enabled by temporally resolving the dynamics better than the sub-optical cycle of the applied laser field, utilizing the method of attosecond transient absorption spectroscopy (ATAS) \cite{Loh2007PRL}. We identify the spectroscopic fingerprint of instantaneous polarization and breaking of the symmetry of the atom in the intense visible femtosecond pulse used in an ATAS experiment and develop a time domain picture describing the dipole response that leads to these signatures. In the future, this general experimental approach can be used to measure strong-field induced symmetry-breaking effects in other atomic or molecular systems also consisting of two or more active electrons and thus to examine new routes to their bound-state and transition-state control by laser fields.
\end{abstract}

\maketitle

The influence of external electric fields on atomic systems can be classified into different regimes of field strength. Weak perturbative field strengths change and shift the energy structure according to the Stark-effect \cite{Autler1955,Stark1914}, whereas for increasing field strengths the symmetry of the atomic potential is broken and the rules governing dynamic processes within the atom change fundamentally. We study the impact of strong electric fields on bound-state dynamics in atoms by examining spectral signatures occurring for increasing laser intensity in the scheme of attosecond transient absorption spectroscopy (ATAS) schematically shown in figure~1a).  
In previous ATAS measurements and theoretical investigations \cite{Chen2012,Chen2012OptLett}, light-induced states (LIS) \cite{Schafer2012PRA} were observed and explained as nonlinear processes involving one or more near-visible (VIS) photons after the initial XUV-excitation step. These LIS can be understood in a dressed state picture using the rotating wave approximation (RWA), where the observed spectral signatures are described as resonant population transfer between states of different symmetries \cite{Chen2012}. In this picture, that is also valid for large time delay values, which-way interferences of different excitation pathways lead to features with characteristic hyperbolic oscillation structures in the recorded absorption spectra \cite{Chini2014JPhysB}. Effects beyond the RWA were investigated in \cite{Pfeiffer2012PRA}, where an analytical equation for the time evolution of a three-level system dressed by a strong femtosecond pulse is obtained. In another work, a sub-cycle AC-Stark effect was identified to be responsible for periodic energy shifts of absorption signatures that belong to dipole-allowed transitions \cite{Chini2012PRL}. In this work we investigate the opposite extreme: a system interacting with a constant (DC) electric field at each point in time during the strong and short laser pulse. We concentrate on the region of pulse overlap where system is dressed by the VIS-pulse during the XUV-excitation step. As we show below, this instantaneous DC-Stark effect allows to understand sub-cycle oscillations of LIS as the consequence of atomic parity-symmetry breaking. We start out with the following question: What is the spectroscopic signature of instantaneous parity-symmetry breaking for atomic bound states in a strong time-dependent field? To this end, we consider the most simple case of a three-level system that consists of the excited states $|o\rangle$, $|e\rangle$ and the ground state $|g\rangle$ with respective energies $E_o$, $E_e$ and $E_g$. Here, $|o\rangle$ possesses odd parity symmetry and is therefore a dipole-allowed transition (transition dipole moment $d_{g,0}\neq 0$) from an even-parity ground state, while $|e\rangle$ has even parity symmetry which means its ground-state transition is dipole forbidden ($d_{g,0}=0$).  We consider the energy separation of both excited states to the ground state much larger (extreme ultraviolet, XUV) than between the excited states (infrared, IR).  The sub-system of states $|o\rangle$ and $|e\rangle$ interacting with a strong DC electric field $\mathcal{E}$ can be described by diagonalizing the Hamiltonian
\begin{align}
H=
\begin{pmatrix}
E_o & \Omega\\
\Omega & E_e
\end{pmatrix}
\end{align}
The interaction is considered in the dipole approximation with $\Omega=d_{o,e}\cdot \mathcal{E}/\hbar$ ($d_{o,e}$ is the transition dipole moment between the states $|o\rangle$ and $|e\rangle$) denoting the Rabi-frequency.
After diagonalization, the states $\left|+\right>$ and $\left|-\right>$ form the new basis set, obtained for the two-level interacting with the DC field $\mathcal{E}$
\begin{align}
\left|+\right>&=\cos(\theta(\mathcal{E}))\left|o\right>+\sin(\theta(\mathcal{E}))\left|e\right>\nonumber\\
\left|-\right>&=\cos(\theta(\mathcal{E}))\left|e\right>-\sin(\theta(\mathcal{E}))\left|o\right>
\end{align}
$\left|+\right>$ and $\left|-\right>$ are a mixture of the excited states $|e\rangle$ and $|o\rangle$ with contributions depending on the field strength of the DC electric field. The mixing angle $\theta$ is defined by $\tan(2\theta)=-\Omega(t)/\Delta E$ \cite{Dalibard1985} and determines the contribution of the original states $|o\rangle$ and $|e\rangle$ to the dressed states. Here $\Omega$ is the Rabi-frequency and $\Delta E$ denotes the energy-level spacing between $|o\rangle$ and $|e\rangle$.

\begin{figure}[h!]
\centerline{\includegraphics[width=7cm]{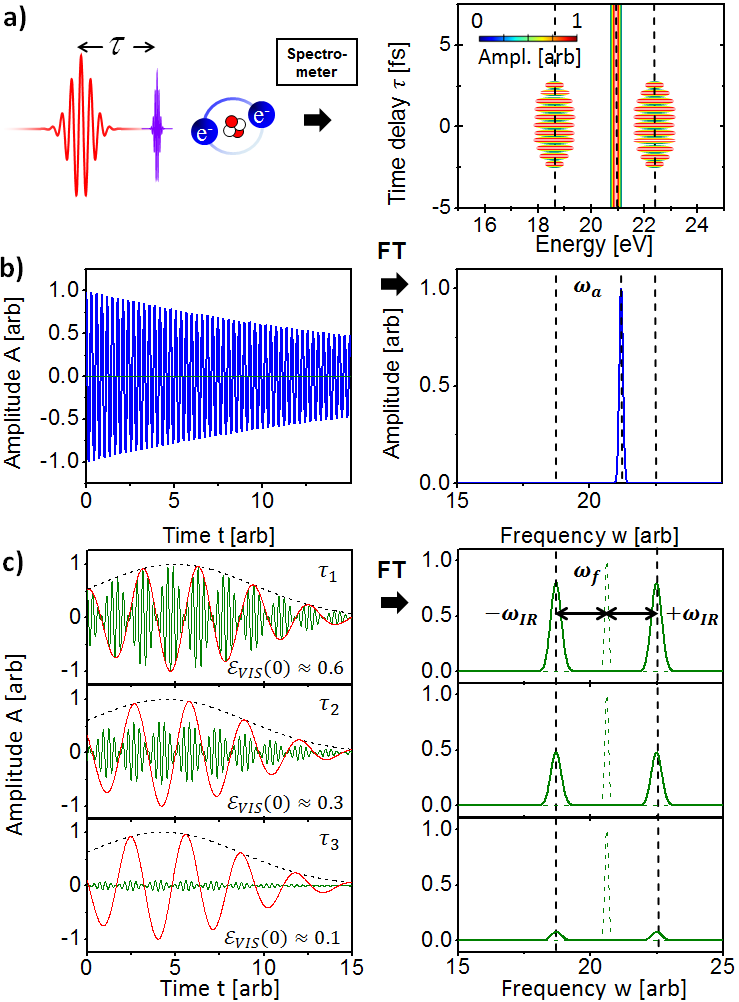}}
\caption{\scriptsize\textbf{Principle of the polarization effect:} \textbf{a):} Illustration of the transient-absorption experiment. A strong VIS laser pulse interacts with the bound states of an atom, at variable time delay $\tau$ after the XUV excitation at $t=0$. Scanning $\tau$ yields the absorption spectrum of a strong-field driven atom. \textbf{b):} Blue line: Unperturbed decaying dipole moment of a dipole-allowed transition at frequency $\omega_a$ with its Fourier-transformation, a Lorentzian line profile. \textbf{c):} Green line: Decaying time-dependent dipole moment of a dipole-forbidden transition at $\omega_f$, enabled by  DC-Stark mixing in the VIS laser field (red line) with its Fourier transform (black dashed line indicates the pulse envelope). The dipole is only active and radiates for a non-zero electric field (due to mixing with the dipole-allowed transition). As a consequence of this beating pattern of the dipole response in the time domain, the Lorentzian lines in the spectrum appear shifted by the carrier frequency of the VIS pulse. The time-dependent dipole moment and its Fourier transform are shown for three different time delays which illustrates the modulation of the amplitude of the signature which depends on the time delay with a $2\omega_{VIS}$ periodicity.  This is due to the fact that the initial excitation of the dipole-forbidden transition by the XUV field (t=0) can only occur for a non-zero VIS field $\mathcal{E}_{VIS}(t=0)\neq 0$}
\end{figure}

We now introduce a time dependence to the Hamiltonian by replacing the DC field $\mathcal{E}$ with the time-dependent field of a few-cycle VIS laser pulse $\mathcal{E}(t)$ which arrives at time delay $\tau$ after the excitation with the XUV pulse at $t=0$. This results in time-dependent energies $E_{\pm}(t)$ and dressed states $|\pm(t)\rangle$ given by the instantaneous electric field $\mathcal{E}(t-\tau)$, and introduces the time-dependent Rabi frequency $\Omega=d_{o,e}\cdot \mathcal{E}(t-\tau)/\hbar$. Furthermore, after an XUV excitation at time $t=0$, both dressed states can be excited by means of the non-zero dipole matrix elements and thus the initial population coefficients $a_+=\left<g|d|+(t)\right>$ and $a_-=\left<g|d|-(t)\right>$ now explicitly depend on the time delay $\tau$, given by the instantaneous electric field $\mathcal{E(-\tau)}$ at the time of excitation by the XUV pulse. The wave function is now given by

\begin{align}
\left|\Psi(t)\right>\propto\phantom{+}&a_+(\mathcal{E}(-\tau))e^{-\imath E_+(t)t}\left|+(t)\right>\nonumber\\
+&a_-(\mathcal{E}(-\tau))e^{-\imath E_-(t)t}\left|-(t)\right>
\end{align}

$E_+$ and $E_-$ are denoting the eigenenergies in this dressed-state picture of a constant electric field \cite{Dalibard1985}. Within this adiabatic approximation the initial population coefficients $a_{\pm}(\mathcal{E}(-\tau))$ are treated as constant after the excitation step which disregards population transfer after the initial population of the excited states in the VIS field. From this ansatz, a rather compact analytical expression for the time-dependent dipole response can be found
\begin{widetext}
\begin{align}
D&(t,\tau)=\notag\\
&d_{g,o}\cdot\left[-2a_+(\tau)\cdot\sin\left(\frac{1}{2}\arctan\left(-\frac{\Omega}{\Delta E}\right)\right)\cdot\cos(E_+(t)t)\right.\notag\\ 
&\left.+2a_-(\tau)\cdot\cos\left(\frac{1}{2}\arctan\left(-\frac{\Omega}{\Delta E}\right)\right)\cdot\cos(E_-(t)t)\right]
\end{align}
\end{widetext}
The discussed treatment in an adiabatic approximation is only valid for moderately strong electric fields with $\Omega(\mathcal{E}(t))<\omega_{VIS}<\Delta E$, where $\omega_{VIS}$ is the central photon energy of the VIS pulse. In the case of $\omega_{VIS}>\Delta E$ the phase of the time-delay dependence of the virtual states resulting from the modulation of the time-dependent dipole moment due to the instantaneous polarization changes. This will be shown below and can be used to discern different coupling cases in the strongly driven system. For the following discussion, we chose the unperturbed state energies to coincide with the $1s2s$ and $1s2p$ states in Helium. The first term of equation~(4) approaches the dipole response of the $1s2s$ state for electric field strength $\mathcal{E}\rightarrow 0$ whereas the second term converges to the contribution of the $1s2p$ state.  

\begin{figure*}[ht]
\centerline{\includegraphics[width=16cm]{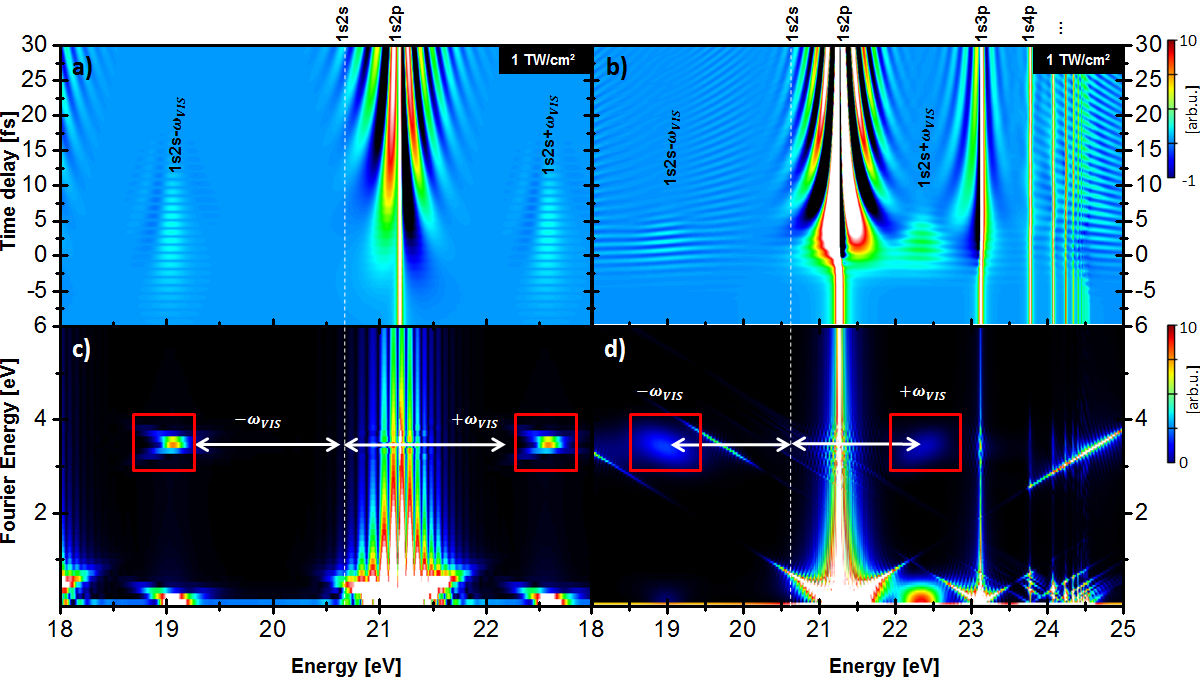}}
\caption{\scriptsize\textbf{Simulation results:} \textbf{a),c):} Time-delay scan and Fourier-evaluation along the time-delay axis of the simulation using equation~(4). White dashed line: position of the $1s2s$ state (dipole forbidden from the ground state). \textbf{b),d):} Time-delay scan and Fourier-evaluation along the time-delay axis of the full numerical few-level Schr{\"o}dinger equation calculation. White dashed line: position of 1s2s state. Red boxes indicate the distinct 2DAS features of the polarization effect which is also present for the full Schr{\"o}dinger equation calculation.}
\end{figure*}

During XUV--VIS pulse temporal overlap, the polarization caused by the strong VIS field breaks the atom's inversion symmetry and mixes states of different parity. As a consequence, transitions are no longer restricted by the dipole-transition rules. This allows for the dressed $2s$ ($\left|-\right>$) state to be excited by the XUV pulse and thus to give rise to spectroscopic signatures. The form of these signatures can be understood by the time-dependent dipole moments stemming from the dressed states $\left|+\right>$ and $\left|-\right>$ in a strong oscillating external electric field that causes periodic state mixing, keeping in mind that these states asymptotically converge to the 2p and the 2s state for $\mathcal{E}_{VIS}\rightarrow 0$. Figure~1b) shows an unperturbed decaying dipole response of the $2p$ state corresponding to a Lorentzian line in the spectrum. Figure~1c) shows the dipole contribution of the $2s$ state, originating from the first term in equation~(4). The strength of the radiation from the $2s$ state is approximately proportional to the electric field strength of the modulating pulse and therefore follows its periodic oscillation which in turn modulates the time-dependent dipole response of the associated transition.

A Fourier-transform of such a modulated decay (beat pattern) as it is shown in figure~1c) yields a double-line feature with the lines shifted by one photon energy of the polarizing pulse. This spectral signature is the physical consequence of atomic symmetry breaking by the instantaneous DC-Stark effect \cite{Stark1914}, following the sub-cycle temporal evolution of the intense laser field. 

The time-delay dependence is shown in figure~2a). The simulation shows signatures appearing in the region of the XUV--VIS pulse overlap. A Fourier transform is performed along the time-delay axis which yields a two-dimensional absorption spectrum (2DAS) \cite{Blaettermann2014}, plotted in figure~2c). The evaluation reveals a distinct peak at $E=2\hbar\omega_{VIS}$ Fourier-energy.  The reason for this is that the exciting XUV pulse scans the electric field of the VIS pulse in time. At a time of VIS field strength $\mathcal{E}(t)=0$ there is no DC-stark mixing and therefore the $2s$ state cannot be excited and radiate. As the absolute value of the electric field has $2\omega_{VIS}$-periodicity in time, these $2s$-spectral features are observed with that same periodicity which yields the observed $2\omega_{VIS}$-oscillation along the time-delay axis. As opposed to the $2\omega$ modulation in which-way interferences, leading to hyperbolic lines of absorption maxima and minima in the transient-absorption trace, the signature of DC-stark mixing is a frequency-independent absorption modulation. This periodicity, observed in the 2DAS Fourier spectrum as a broad (blob-like) feature provides a unique signature to pinpoint these strong-field symmetry-breaking features explained by the DC Stark effect.\\

To test the validity of the simple adiabatic two-level coupling approach, we compare it to a fully dynamical few-level Schr{\"o}dinger equation model. Within the single-active electron approximation, we here consider only singly-excited states that include the $1s2s$ state and the $1snp$ series up to $n=12$.
The simulated time-delay scan in this model is shown in figure~2b) along with the 2DAS Fourier spectrum.  Again, the DC-Stark signatures can be found at the expected positions: At the energies $E=\omega_{VIS}$ above and below the energy of the $1s2s$ state, and at $2\omega_{VIS}$ Fourier energy. Compared to the DC-Stark feature, identified by the horizontal modulations across the whole spectrum, the which-way interference contributions are in general tilted except near $\tau$=0. This in turn translates to a broad signature of the instantaneous DC-Stark effect in the 2DAS-plot which is clearly distinguishable from the sharp diagonal features understood as which-way interferences \cite{Chini2014JPhysB}.

\begin{figure*}[ht!]
\centerline{\includegraphics[width=16cm]{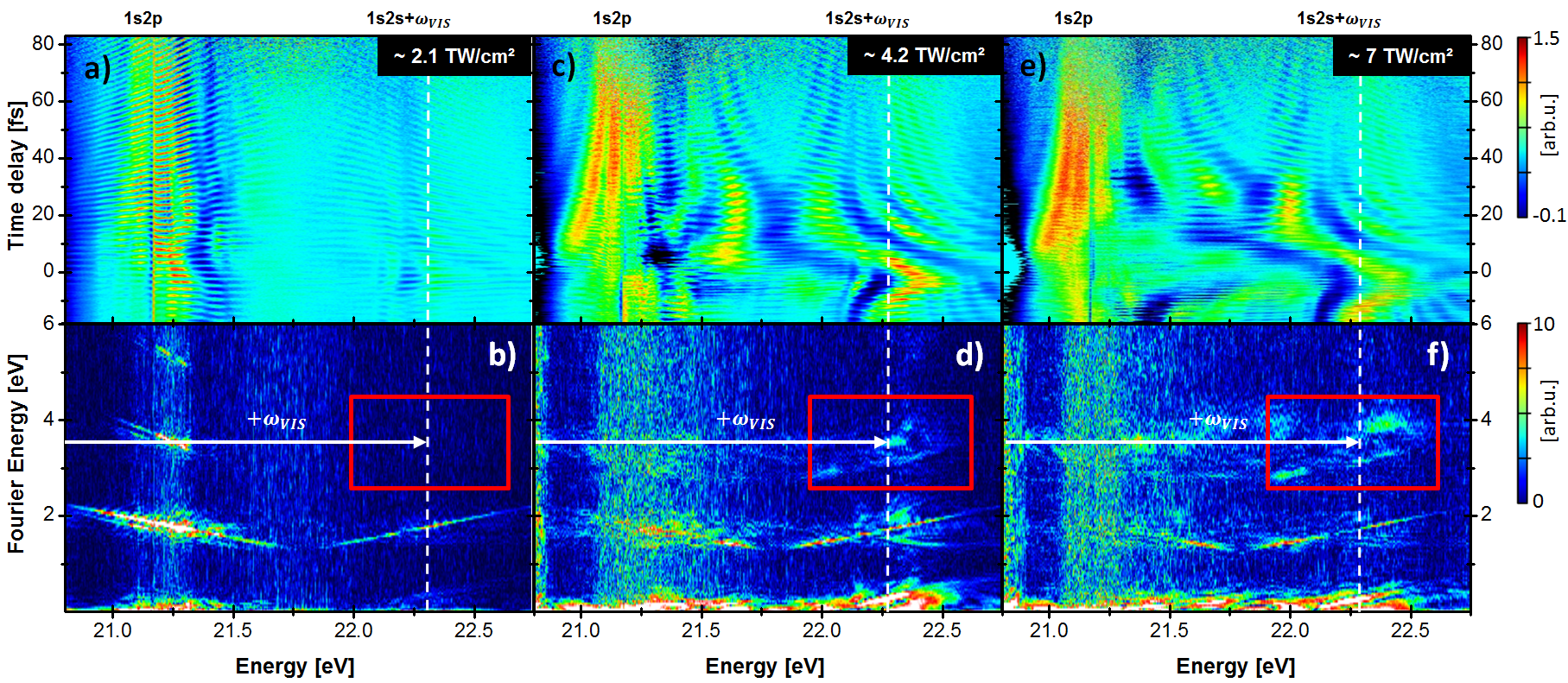}}
\caption{\scriptsize\textbf{Measured XUV-Photon transmission spectra with Helium as target gas:} \textbf{a),c),e):} Time-delay scans for different intensities as given in the plot, LIS appear at $22.3$ eV. \textbf{b),d),f):} Fourier-transformation along the time-delay axis reveals the distinct broad 2DAS-feature of the DC-Stark dressed states, marked by a red box.}
\end{figure*}

An experiment on atomic Helium was performed to observe the predicted effects (see Methods for setup details). Figures~3a), c) and e) show experimentally recorded time-delay scans exhibiting tilted modulation features for the dipole-allowed transitions related to coupling between different states \cite{Ott2012Arxiv}, as well as the emerging LIS between the excited states of Helium \cite{Chini2013SciRep,Bell2013JModOpt}. As expected from the theoretical calculations light-induced absorption features appear at $\sim$ 22.3 eV which is one VIS photon energy ($\sim$ 1.7 eV) above the $2s$ state at 20.62 eV and one photon energy below the $5s$ state at 24.01 eV. The 2DAS trace is used to obtain the plots in Figure~3b), d) and f). Marked in the red boxes, they show the expected DC-Stark signature: a broad distribution in contrast to the diagonal coupling features and located at one photon energy above the transition energy of the $2s$ state and at the Fourier-energy of $2\omega_{VIS}=3.4$ eV.  Due to the limited spectral range of our spectroscopic grating the lower-lying part of the spectroscopic signature of the $1s2s$ state at $\sim$ 18.9 eV could not be measured. It is interesting to note that the features become spectrally broader for higher intensities. This could be tentatively explained by a sub-cycle energy-chirp of the excited-state's dipole response caused by the DC-Stark shift due to the time-dependent strong electric field strength during one half-cycle of the VIS-pulse.

In order to measure the DC-Stark effects as a function of intensity, the characteristic spectral signatures indicated by the red boxes in the 2DAS spectrum (figure~3) are selected by a 2D-Supergaussian filter and then Fourier-transformed back into the time-delay domain. This isolates the time-resolved spectral contribution of the dressed state that originates from the 1s2s state. The results are given in figure~4a) to f). As these plots clearly show, the DC-Stark-induced dressed states emerge as theoretically expected for increasing intensity at $E_{1s2s}+\omega_{VIS}$ at time-delays where the XUV- and VIS-pulses overlap and are strongest for the exact pulse overlap, at time-delay $\tau=0$. In accordance with theory the signatures become stronger for increasing intensity. 

\begin{figure}[ht!]
\centerline{\includegraphics[width=8cm]{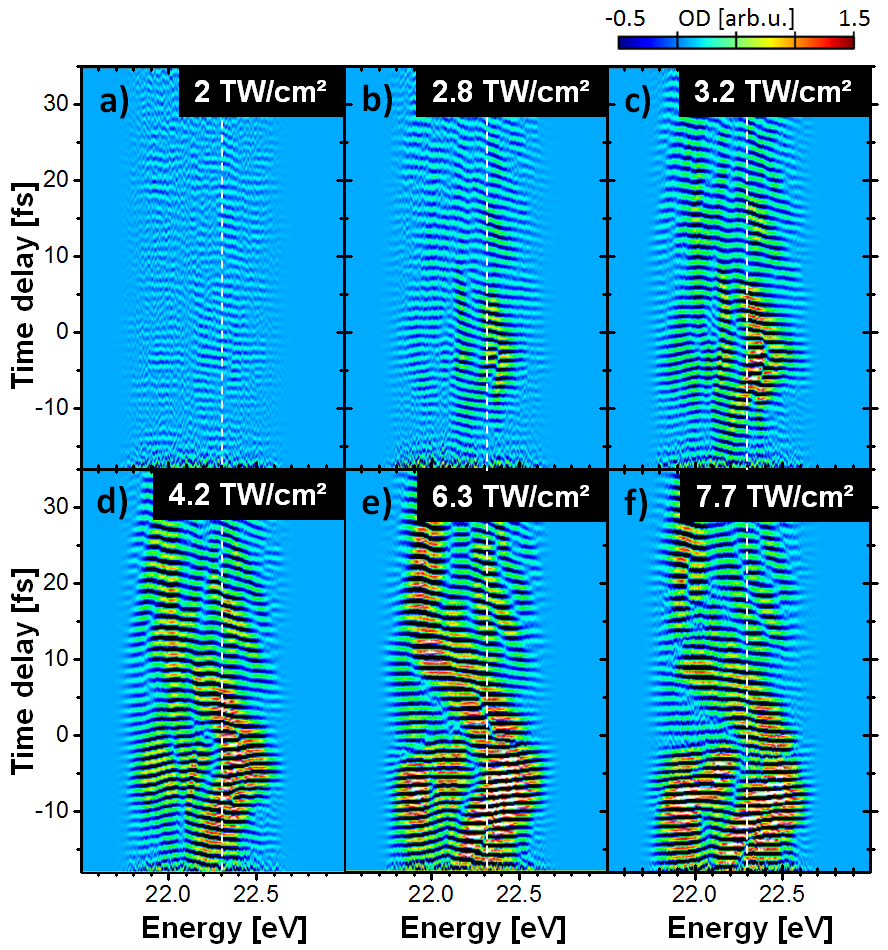}}
\caption{\scriptsize\textbf{Extracted signatures of the sub-cycle polarization effect:} \textbf{a)} to \textbf{f)}: Isolation of the time-delay dependent absorption of the DC-Stark modulated $2s$ state by Fourier transform of the corresponding LIS signature in the experimental data (red boxes in figure~3). At low intensities, the DC-Stark dressed state only contributes during XUV/VIS pulse overlap near time delay $\tau=0$. For higher intensities, also weaker parts of the VIS pulse at earlier or later times (pedestal) are sufficient to create the absorption response. Besides the dominant 2s dressed state response (marked by white dashed line), also several higher-lying $ns$ or $nd$ states seem to contribute in this energy range.}
\end{figure}

Interestingly the energy of the observed signatures does not shift to lower values as expected for an isolated 1s2s DC-stark shift but goes up during the pulse overlap. Also there are indications of multiple absorption features appearing around 22.3 eV. A tentative explanation for this could be DC-Stark contributions of states other than just the $1s2s$ state, specifically the $1s5s$, the $1s5d$, or even higher-lying states within the Rydberg manifold, approaching the first ionization threshold of helium at $24.6$ eV. During the interactions with the VIS-pulse these states are near-degenerate and thus energetically repel each other in the energy-domain when coupled by the intense VIS pulse. The observed modulations also show different phases with respect to each other which points towards
different polarization effects if for example $\Delta E$ becomes smaller than the VIS photon energy like in the case of the coupling between the 1s2s and 1s2p states.
A full understanding of the details of this complicated strong-field induced dynamics goes beyond the scope of this letter and can be used to test and benchmark quantum-dynamics theories in the future.
We expect the methodology demonstrated here can also be applied to the study of strong-field symmetry breaking in other target species such as poly-electronic atoms or molecules.\\
\\

\textbf{Methods}
\\

The experimental setup used for the measurements consists of a commercially available Ti:Sapphire multi-pass amplifier laser system (Femtolasers Compact Pro) including a hollow-core fiber with a subsequent chirped-mirror compression stage for the generation of near-visible (VIS) laser pulses of 5-7 fs duration, photon energy of $\sim$1.7 eV, $\sim350$ $\mu$J peak intensity and 4 kHz repetition rate.
These pulses are focused into a gas cell filled with xenon ($\sim$20 mbar backing pressure), yielding peak intensities in the $10^{13}$ W/cm$^2$ regime, and act as the driving field for the high-harmonic generation (HHG) process which creates broadband attosecond extreme ultraviolet (XUV) pulses phase locked to the driving VIS field. Both XUV and VIS pulses are focused into the experimental target, a cell filled with helium gas ($\sim$60 mbar backing pressure), where the broadband XUV pulse, with photon energy ranging from 18 eV up to 30 eV which covers the spectrum of singly excited helium, serves as a coherent excitation pulse in the presence of the strong few-cycle VIS electric field.\\
The control parameters of the experiment are the time delay between XUV and VIS pulse and the intensity of the VIS pulse. The relative time delay is controlled with a piezo-driven two-component split-mirror setup (10 as precision) which temporally separates the co-propagating pulses. The intensity of the VIS pulses after HHG is simultaneously controlled by a piezo-driven iris aperture. In order to get almost full spatial and temporal separation of the XUV and VIS beams a two-component band-pass filter in annular geometry is used (center part: 0.2 $\mu$m aluminum filter, outer part: 2 $\mu$m nitrocellulose filter).\\
The transmitted XUV radiation which contains information about the dipole response of the system is recorded by a high-resolution home-built XUV spectrometer (resolution: 10 meV FWHM at 20 eV). The experimental and spectroscopic setup is placed in vacuum chambers and operated at pressures of $10^{-3}$ mbar. The acquired data was analyzed by using a Fourier filter to reconstruct an in-situ reference spectrum~\cite{Ott2013}.\\


\end{document}